# Diffuse fraction as a tool for exploring the sensitivity of parametric clear-sky models to changing aerosol conditions

Robert Blaga[1], Oana Mares[2], Eugenia Paulescu[1,*], Remus Boata[3], Andreea Sabadus[4], Sergiu-Mihai Hategan[1,4], Delia Calinoiu[2], Nicoleta Stefu[1], Marius Paulescu[1]

[1]Faculty of Physics, West University of Timisoara, V. Pârvan 4, 300223 Timișoara, Romania

[2]Department of Fundamental of Physics for Engineers, Politehnica University Timișoara, V. Pârvan 2, 300223 Timisoara, Romania

[3]Astronomical Institute of the Romanian Academy, Timisoara Astronomical Observatory, A. Sever Sq. 1, 300210, Timisoara, Romania

[4]Institute for Advanced Environmental Research, West University of Timișoara, V. Pârvan 4, 300223, Timișoara, Romania

Corresponding author: eugenia.paulescu@e-uvt.ro

ABSTRACT

Aerosols' impact on the performance of a clear-sky solar irradiance model is often evaluated from the perspective of the overall accuracy of estimates. This study assesses the aerosol role in clear-sky solar irradiance modelling from a totally different perspective, namely the ability of a model to accurately separate global solar irradiance into its fundamental direct and diffuse components. In an innovative approach, the analysis is focused on the sensitivity of parametric solar irradiance models to changes in aerosol conditions. A new measure, the aerosol influence quantifier $\Omega$, is introduced for assessing the correlation strength between the relative variation in clear-sky diffuse fraction and the relative variation in atmospheric aerosol load. The effectiveness of the aerosol influence quantifier is explored as a tool for assessing the sensitivity of three parametric clear-sky models to variations in input aerosol data. A rather surprising result is reported, e.g. if high quality aerosol data are not available, one could potentially maximize the estimate's accuracy by using a parametric clear-sky solar irradiance model with low sensitivity to aerosol variations. In such cases, it is shown that the aerosol influence quantifier $\Omega$ can be used as a tool for identifying an adequate model for input data with a given uncertainty.

KEYWORDS: solar irradiance; clear-sky; diffuse fraction; aerosol; model sensitivity;

List of symbols

| Symbol | |
|---|---|
| AOD | Aerosol optical depth |
| ETR | Extraterrestrial radiation |
| $g$ | Aerosol asymmetry factor |
| $G$ | Global horizontal irradiance |
| $G_b$ | Beam horizontal irradiance |
| $G_d$ | Diffuse horizontal irradiance |
| $k_d$ | Diffuse fraction |
| $k_{d,cs}$ | Diffuse fraction under clear sky |
| $l_{O3}$ | Ozone column content |
| $l_{NO2}$ | Nitrogen dioxide column content |
| $m$ | Atmospheric optical mass |
| $p$ | Atmospheric pressure |
| $w$ | Water vapor column content |
| $\alpha$ | Ångström exponent |
| $\beta$ | Ångström turbidity coefficient |
| $\gamma$ | Downward fraction of scattered radiation |
| $\rho$ | Ground albedo |
| $\tau_a$ | Specific atmospheric transmittance associated with aerosol scattering |
| $\tau_R$ | Specific atmospheric transmittance associated with Rayleigh scattering |
| $\varpi$ | Single scattering albedo |
| $\Omega$ | Aerosol influence quantifier |

1. INTRODUCTION

At a generic level, the estimation of solar energy over a given period can be viewed as a two-stage procedure. Firstly, the amount of collectable solar energy under clear sky is evaluated [1]. Secondly, the result is adapted to the current state-of-the-sky, for instance, by using an Ångström equation [2]. In such a two-stage process, the main source of uncertainty is commonly associated with the inadequate description of clouds transmittance. However, this uncertainty may be equally caused by weaknesses of the clear-sky solar irradiance models in the evaluation of the atmospheric transmittance. Different models estimate the atmospheric transmittance at different levels of accuracy, achieving different performances in clear-sky solar irradiance estimation [3].

The extraterrestrial solar radiation flux is changed by the atmospheric constituents. A large alteration is due to aerosols (dust, soot, organic carbon, sea salt, etc.) through scattering and, to a lesser extent, absorption. To complete the picture, aerosols are by far the most unstable atmospheric constituent. The substantial effect of aerosols on the collectable solar energy under clear sky can be gauged, for example, from [4]. Taking the climatological aerosol environment as reference, the study demonstrated that during erratic episodes with high atmospheric aerosol loading, the collectable solar energy can decrease by more than 20% in a temperate continental climate.

In recent years, the networks that measure the properties of aerosols on the ground, such as the Aerosol Robotic Network – AERONET [5], have been intensively developed. Aerosols properties evaluated from reanalysis products, such as MERRA-2 [6] or retrieved from satellite observation, such as MODIS [7], become more and more popular inputs for solar irradiance models. However, many databases, accessed for retrieving the aerosol parameters required for running a clear sky solar irradiance model with global coverage, store the aerosol parameters with monthly to yearly sampling [8]. Solar irradiance estimated using monthly-averaged aerosol optical depth (AOD) was found biased compared to solar irradiance estimated using daily AOD [9]. Improving the aerosol representation in a clear-sky solar irradiance model may increase the overall model performance. This statement applies to the numerical weather prediction models too [10]. Ideally, solar irradiance estimation through clear-sky models should be done with aerosol data no coarser than hourly, to capture intra-day variations in aerosol loading.

The temporal resolution of aerosol properties may have a major impact on the solar irradiance estimates. Most solar irradiance models account for aerosol effect on the atmospheric transmittance in terms of the Ångström turbidity coefficient [11]. The Ångström turbidity coefficient $\beta$ is defined as the AOD measured at the wavelength of 1 μm. The popularity of the Ångström turbidity coefficient originates in its intrinsic meaning, namely a measure of the atmospheric aerosol loading. $β$ typically ranges between 0 (no aerosols) and 0.5 (high aerosol load). A value tending to one or higher indicates an extremely polluted atmosphere (e.g. a haze event). The relative error in estimating global solar irradiance under clear-sky due to an inadequate consideration of the Ångström turbidity coefficient may be considerable. For instance, when the current value of the Ångström turbidity coefficient is replaced by the yearly average value, the relative errors in estimating global solar irradiance may be higher than 20% [12].

There are some studies focused on quantifying the effect of aerosols' variability on solar irradiance. For example, Ref. [13] proposes two measures for solar irradiance variability under clear sky induced by AOD: the Aerosol Variability Index (AVI) which characterizes the magnitude of the variability in AOD, and the Aerosol Sensitivity Index (ASI) which relates the magnitude of relative variations in solar irradiance to absolute variations in AOD. The aerosol-induced variability in global solar irradiance was found 2–4 times lower than that in direct normal irradiance [13].

By far the most studies are devoted to the evaluation of aerosol influence on clear-sky solar irradiance. Very few studies focus on the aerosol impact on the separation of clear-sky global solar irradiance into its primary components direct-normal and diffuse. In a recent study, the classical diffuse fraction was explored as an appropriate quantifier for the fractional part of the global solar irradiance estimated by a clear-sky solar irradiance model as being diffuse [14]. It was shown that the diffuse fraction is an adequate tool for isolating the uncertainty induced by aerosols in estimating the diffuse solar irradiance under clear sky. Testing the accuracy of a model in estimating the clear-sky diffuse fraction completes the standard validation of a clear-sky solar irradiance model, by statistical comparison of the estimates with measurements. A strong influence of the atmospheric aerosol loading on the dissimilarities between the accuracies in estimating clear-sky diffuse fraction and the related diffuse solar irradiance was noticed.

This work continues the study [14], revealing new capabilities of the clear-sky diffuse fraction as a quantifier of the aerosol impact on the global solar irradiance. In Ref. [14] the estimation errors are separated into two classes: (A) intrinsic/general model errors and (B) aerosol-related errors. The aerosol-related errors may be further divided into two sub-classes: (B1) the inability of the solar irradiance model in capturing the aerosol influence on the atmospheric transmittance and (B2) the uncertainty in the input parameters associated with aerosols. Topic B2 was not addressed at all in Ref. [14]. Thus, in this study we build on our work [14], but from a very different perspective: we study the sensitivity of clear-sky solar irradiance models to the variations in the aerosol properties. Sensitivity to aerosols is a basic characteristic of each clear-sky solar irradiance model. However, the sensitivity to aerosols can indirectly quantify the impact of the aerosol data uncertainty on the models' estimates. To evaluate quantitatively the models' sensitivity a new quantifier for the correlation strength between the relative variation in diffuse fraction and the relative variation in atmospheric aerosol load is introduced. The atmospheric aerosol load is quantified through the Ångström turbidity factor. The new quantifier is employed to study the sensitivity of three different parametric clear-sky solar irradiance models to variations in the aerosol atmospheric loading.

If the aerosol data are accompanied by a high uncertainty (e.g. using as input yearly average values) and the model sensitivity is high, then the aerosol impact on estimates accuracy is high (the errors from B2 sub-class are large). If the aerosol data are accompanied by a high uncertainty and the model sensitivity is low, then the impact is low (the errors from B2 sub-class are small). These two statements are true irrespective of the model ability in capturing the aerosols influence on the atmospheric transmittance (the magnitude of errors in the sub-class B1). In general, it is expected that a model with low sensitivity to be one that threats aerosols in a simplistic way (errors in class B1 will be large) and vice versa. These are intuitive cases, but there are cases that cannot be categorized as intuitive. As shown in the paper, intuitive cases validate the proposed quantifier, while its usefulness is proven in non-intuitive cases.

The paper is organized as follows. Section 2 introduces the new quantifier. In Sec. 3, the new quantifier is employed to map the sensitivity of three representative parametric clear-sky solar irradiance models to changes in aerosol conditions. Section 4 summarizes the main conclusions. For each clear-sky solar irradiance model, mathematical equations of the new quantifier are illustrated in Appendix and their inference is summarized in the Electronic Supplementary Material.

## 2. NEW QUANTIFIERS FOR THE AEROSOL INFLUENCE ON SOLAR RESOURCES

Earth's atmosphere separates the extraterrestrial solar radiation (ETR) flux into two parts: the beam component, which directly reaches the Earth's surface, and the diffuse component, which is the result of the downward scattering of the ETR by the atmosphere. Horizontal beam ($G_b$) and diffuse ($G_d$) solar irradiances are the radiometric quantities associated with these components. The global horizontal solar irradiance ($G$) is defined as the sum of beam and diffuse components:

$$G = G_b + G_d \tag{1}$$

Under clear sky $G_b$ and $G_d$ are frequently modeled with respect to atmospheric mass and several common atmospheric parameters (e.g. water vapor column content, ozone column content, etc.).

This study focuses on the impact of aerosols on the solar irradiance reaching the ground. This impact is complex due to the varying optical properties, chemical composition, interactions and spatio-temporal distribution of the aerosols. As such, the aerosol impacts are hard to model in detail. Overall, up to moderate atmospheric aerosol loading, the increase in aerosols mainly causes an increase in the diffuse component and a decrease of the beam component. This process takes place without energy conservation, a fraction being lost by upward scattering and aerosol absorption. The increasing of the atmospheric aerosol load causes not only a change in the diffuse to beam irradiance ratio, but also a reduction of the global solar irradiance. Thus, for a multifaceted characterization of the influence of aerosols on solar irradiance, a quantifier for evaluating the changes in the diffuse to beam irradiance ratio $G_d/G_b$ is useful.

## 2.1. Clear-sky diffuse fraction ($k_{d,cs}$)

The diffuse fraction is defined as the ratio between the diffuse $G_d$ and global $G$ solar irradiances at the ground level:

$$k_d \equiv \frac{G_d}{G} = 1 - \frac{1}{1 + G_d/G_b} \tag{2}$$

The diffuse fraction is a radiometric parameter routinely used in solar energy modeling under all-sky conditions. Starting with the reference work of Liu and Jordan [15], many authors have dealt with modeling $k_d$. The main goal of these studies was the development of an empirical relationship between diffuse fraction and clearness index (see e.g. [16] and the discussion therein). For such models, the global solar irradiance is assumed to be measured, and is used to generate estimates for the diffuse and beam components.

This study looks at diffuse fraction from a very different perspective. Under clear sky conditions, $k_d$ primarily depends on the atmospheric aerosol loading and the air mass $m$ (see Sec.

3.2 for a discussion). Thus, the clear-sky diffuse fraction $k_{d,cs}$ appears naturally as a quantifier for the way in which the atmosphere without clouds separates the global solar irradiance into beam and diffuse components through aerosol absorption and scattering, and Rayleigh scattering. A preliminary analysis of $k_{d,cs}$ as quantifier for the global to beam/diffuse partition was explored in [14].

### 2.2 Aerosol influence quantifier

At a generic level, the relationship between the variation of clear-sky diffuse fraction $k_{d,cs}$ and the variation of the Ångström turbidity coefficient $\beta$, in relative terms, reads:

$$\frac{dk_{d,cs}}{k_{d,cs}} = \Omega(m,\beta)\frac{d\beta}{\beta} \qquad (3)$$

With Eq. (3), we introduce a new quantifier $\Omega$, which captures the effect of changes in the aerosol amount on the clear-sky diffuse fraction. The aerosol influence quantifier $\Omega$ primarily depends on the atmospheric air mass and the Ångström turbidity coefficient. Dependence of the solar irradiance on the finer aerosol properties (e.g. single scattering albedo or aerosol asymmetry factor) is also captured.

When evaluated on measured data, $\Omega$ gives information about the sensitivity of the incoming solar irradiance on the aerosol loading and radiative properties. However, through empirical exploration, we have found that the implementation of $\Omega$ exclusively on measured data is difficult. The differentials in Eq. (3) must be evaluated while keeping all the other influential parameters, besides $\beta$, constant. In principle, this means that one could take two data points (diffuse fraction) that are not from the same time/location, but measured in the same atmospheric conditions (the same values for all atmospheric parameters are recorded, including the atmospheric air mass) and evaluate $\Omega$. In practice, as Appendix A illustrates, the broadband description of solar irradiance proves too coarse to obtain meaningful results through this operation.

If $k_{d,cs}$ is modelled instead of measured, $\Omega$ quantifies the sensitivity of the underlying clear-sky model to changes (or uncertainties) in the aerosol data used as input. Indeed, given a physical clear-sky model, the aerosol influence quantifier $\Omega$ can be easily implemented either through finite difference method or using its analytical expression, as discussed in the following Section. .

## 3. EVALUATION OF AEROSOL INFLUENCE QUANTIFIER THROUGH CLEAR-SKY SOLAR IRRADIANCE MODELS

In this section, equations for the aerosol influence quantifier $\Omega$ are deduced, considering three representative parametric clear-sky solar irradiance models (one very simple, another optimized for the estimation of the diffuse component and the third very performant). The sensitivity of the models is mapped under the range of atmospheric parameters collected from ground measurements. The utility of $\Omega$, as a tool for measuring the sensitivity of parametric clear-sky models to changes in the input aerosol data is explored.

### *3.1. Working database*

The properties of the clear-sky diffuse fraction $k_{d,cs}$ measured by the aerosol influence quantifier $\Omega$ are analytically explored, but using atmospheric and radiometric recorded parameters. For this a database was compiled from two sources: Aerosol Robotic Network AERONET [17] and Baseline Surface Radiation Network BSRN [18]. Aerosol optical properties and other relevant atmospheric parameters were retrieved from AERONET, while radiometric data were retrieved from BSRN. Table 1 presents the 12 stations considered for this study, their local climate according to the Köppen-Geiger classification [19], and the number of records. The stations were chosen based on a single principle: a BSRN station to be near an AERONET station.

TABLE 1. Summary of the stations from where the data were collected, stratified according to the Köppen-Geiger climate classification [19]. *N* represents the number of clear-sky records recorded from a station.

| Nr | Country | Location | Climate | BSRN index | BSRN station | | | AERONET station | | | N |
|----|---------|----------|---------|------------|--------------|---|---|-----------------|---|---|---|
|    |         |          |         |            | Lat. [deg] | Long. [deg] | Alt. [m] | Lat. [deg] | Long. [deg] | Alt. [m] |   |
| 1  | Nigeria | Ilorin | Aw | ILO | 8.53 | 4.56 | 350 | 8.32 | 4.34 | 350 | 89 |
| 2  | Algeria | Tamanrasset | Bwh | TAM | 22.79 | 5.52 | 1385 | 22.79 | 5.53 | 1377 | 548 |
| 3  | Israel | Sede Boqer | Bwh | SBO | 30.85 | 34.77 | 500 | 30.85 | 34.78 | 480 | 548 |
| 4  | Saudi Arabia | Solar Village | Bwh | SOV | 24.91 | 46.41 | 650 | 24.90 | 46.39 | 764 | 424 |
| 5  | USA | Tucson | Bwh | TUC | 32.22 | -110.95 | 786 | 32.23 | -110.95 | 779 | 369 |
| 6  | Brazil | Petrolina | Bwk | PTR | -9.06 | -40.31 | 387 | -9.38 | -40.50 | 370 | 287 |
| 7  | Brasil | Sao Martinho da Serra | Cfa | SMS | -29.44 | -53.82 | 489 | -29.44 | -53.82 | 489 | 325 |
| 8  | USA | Billings | Cfa | BIL | 36.60 | -97.51 | 317 | 36.60 | -97.48 | 318 | 221 |
| 9  | USA | Boulder | Cfb | BOU | 40.05 | -105.00 | 1577 | 40.05 | -105.00 | 1604 | 394 |
| 10 | France | Palaiseau | Cfb | PAL | 48.71 | 2.20 | 156 | 48.70 | 2.20 | 156 | 584 |
| 11 | France | Carpentras | Csa | CAR | 44.08 | 5.05 | 100 | 44.08 | 5.05 | 100 | 368 |
| 12 | Estonia | Toravere | Dfb | TOR | 58.25 | 26.46 | 70 | 58.25 | 26.46 | 70 | 363 |

TABLE 2. Summary statistics of all columns of the working database. The column meaning is: $p$ – atmospheric pressure; $l_{O3}$ – ozone column content; $l_{NO2}$ – nitrogen dioxide column content; $w$ – water vapor column content; $\beta$ – Ångström turbidity coefficient; $\alpha$ – Ångström exponent; $\varpi$ – single scattering albedo; $g$ – aerosol asymmetry factor; $G$ – global horizontal irradiance; $G_d$ – global horizontal diffuse irradiance. Column content for the atmospheric species is taken from the AERONET database [17], while the solar irradiance quantities are taken from the BSRN database [18].

| Value | Sun elevation [deg] | Altitude [m] | $p$ [hPa] | $l_{O3}$ [DU] | $l_{NO2}$ [DU] | $w$ [g/cm²] | $\beta$ | $\alpha$ | $\varpi$ | $g$ | $G$ [W/m²] | $G_d$ [W/m²] |
|---|---|---|---|---|---|---|---|---|---|---|---|---|
| Minimum | 11.35 | 70 | 835.6 | 251.3 | 0.0593 | 0.2238 | 0.0062 | -0.030 | 0.407 | 0.6167 | 88 | 31 |
| 1st quartile | 20.79 | 156 | 924.2 | 283.5 | 0.1293 | 0.8673 | 0.0252 | 0.5705 | 0.856 | 0.674 | 302 | 64 |
| Median | 37.98 | 500 | 957.8 | 292.9 | 0.1766 | 1.257 | 0.0473 | 1.0893 | 0.894 | 0.699 | 607 | 83 |
| Mean | 39.01 | 605 | 947.7 | 305 | 0.2004 | 1.3673 | 0.1096 | 0.9809 | 0.8743 | 0.7024 | 581.7 | 102 |
| 3rd quartile | 54.44 | 779 | 1000.6 | 320 | 0.2238 | 1.793 | 0.1176 | 1.3515 | 0.924 | 0.725 | 825 | 117 |
| Maximum | 87.83 | 1577 | 1018.3 | 375.3 | 0.3945 | 3.23 | 1.65 | 1.9943 | 0.98 | 0.791 | 1222 | 484 |

Only data recorded in perfect clear-sky days were considered, selected according to the procedure from [14]. Due to the different temporal resolution of the two networks (1-min in BSRN and 15-min in AERONET), in a given day only data recorded simultaneously by the BSRN and AERONET instruments were retained (50 – 60 values per day, depending on the daylight length). Instantaneous values are considered here, i.e. no averaging is applied. The database comprises 4520 entries, each line containing the following simultaneously measured parameters: global horizontal irradiance $G$, diffuse horizontal irradiance $G_d$, atmospheric pressure $p$, ozone column content $l_{O3}$, nitrogen dioxide column content $l_{NO2}$, water vapor column content $w$, Ångström exponent $\alpha$, Ångström turbidity coefficient $\beta$, single scattering albedo $\varpi$, aerosol asymmetry factor $g$, ground albedo $\rho$. A summary statistic over the database is presented in Table 2, highlighting its complexity and diversity. The atmospheric parameters are clustered around values which define normal/climatological values, but many records outside normal atmospheric conditions (even extremes) are included. For instance, roughly half of data are recorded in a moderate loading of the atmosphere with aerosols, but data are also recorded in the conditions of a clean atmosphere (1st quartile of Ångström turbidity coefficient $\beta$ = 0.0252) and, at the other extreme, when the atmosphere is darkened by aerosols ($\beta_{max}$ = 1.65). To conclude, Table 2 shows a challenging nature of the database, including instances from normal atmospheric conditions to extreme atmospheric conditions.

*3.2 Derivation of the aerosol influence quantifier equations for clear-sky models*

Three parametric clear-sky solar irradiance models have been chosen to illustrate the features of the aerosol influence quantifier $\Omega$. Table 3 lists the models, emphasizing their input parameters and the value intervals of the Ångström coefficients for which the models were originally developed.

TABLE 3. List of the models considered in this study with a highlight on the input parameters. See Table 2 for the explanation of the columns. In practice, the local atmospheric pressure is estimated from the site altitude.

| Model | Reference | $p$ | $l_{O3}$ | $l_{NO2}$ | $w$ | $\alpha$ | $\beta$ | $\varpi$ | $g$ | $\rho$ | Observations |
|---|---|---|---|---|---|---|---|---|---|---|---|
| PS | [20] | * | * |  | * |  | * |  |  |  | $\gamma = 0.432$, $\beta < 0.4$ |
| SIMv.2 | [22] | * | * | * | * |  | * | * | * |  | No restriction for $\beta$ |
| REST2 | [24] | * | * | * | * | * | * | * |  | * | $0 < \beta < 1.1$ |
|  |  |  |  |  |  |  |  |  |  |  | $0 < \alpha < 2.5$ |

The first two models (PS and SIMv.2) express at two different levels of complexity the concept of parametric modeling of the clear-sky solar irradiance. PS [20] is a very simple parametric model rooted in the spectral atmospheric transmittances of low resolution of the Leckner's model [21]. SIMv.2 [22] is constructed by weighted-averaging of the spectral atmospheric transmittances from SMARTS2 [23]. In the frame of this study, SIMv.2 is differentiated from PS by the degree of detail of the aerosol properties at input (see Table 3). The REST2 model [24] is structurally different from the previous two models. REST2 was developed using transmittance parameterization based on the SMARTS2 spectral model, but separately for two broad bands of the shortwave spectrum: the UV–VIS band (0.29 – 0.70 μm) and the near-infrared band (0.70 – 4.0 μm). The formalism used to model the diffuse irradiance is based on a two-layer scattering scheme [24]. REST2 is generally accepted as one of the most accurate and robust clear-sky models [3]. REST2 was run neglecting ground reflection. The equations of these models, as they were implemented for this study, are listed in the Electronic Supplementary Material.

Conditioned by their structural difference -but not necessarily determined by it- the three clear-sky models also have different performances. Ref. [3] includes the three models considered here in a list of 95 clear-sky models to be studied according to their performance in estimating direct and diffuse solar irradiance. A ranking of the models was performed using principal component analysis. The overall ranking of the PS, SIMv.2, and REST2 models in estimating the diffuse irradiance was found to be 68, 7, and 4. Their respective ranking for estimating direct irradiance is 52, 38, and 2. We observe that the simpler PS model has modest performance overall. SIMv.2 is modest for direct irradiance, but very accurate for the diffuse component (even achieving 1[st] place in temperate climate). REST2 -in the v5 version implemented here- is very accurate for both diffuse and direct components, which confirms its position as benchmark in the field.

In the case of clear sky models, $\Omega$ is directly expressible by differentiating $k_{d,cs}$ with respect to $\beta$. For each clear-sky solar irradiance model, the computation leads to a different mathematical equation for $\Omega$. The derivation is illustrated next for PS, the simplest of the three models. The expression of $k_{d,cs}$ based on the PS model's atmospheric transmittances reads as:

$$k_{d,cs}(m,\beta) = 1 - \frac{\tau_R \tau_a}{\tau_R \tau_a + \gamma(1 - \tau_R \tau_a)} \quad (4)$$

$\tau_R$ represents the specific atmospheric transmittance associate to the Rayleigh scattering and $\tau_a$ represents the specific atmospheric transmittance associate to the aerosol extinction process in the PS model [20]. $\gamma$ denotes the downward scattering fraction.

By differentiating with respect to $\beta$, after a simple calculus, the following equation for $\Omega$ is obtained:

$$\Omega(m,\beta) = -\beta \frac{\tau_R}{(1-\tau_R \tau_a)\left[\gamma + \tau_R \tau_a (1-\gamma)\right]} \frac{\partial \tau_a}{\partial \beta} \quad (5)$$

PS models the scattering sector only through dependence on air mass and Ångström turbidity, i.e. $\Omega = \Omega(m,\beta)$. For a pure Rayleigh atmosphere ($\gamma = 0.5$), Eq. (5) reduces to:

$$\Omega(m,\beta) = -2\beta \frac{\tau_R}{1-(\tau_R \tau_a)^2} \frac{\partial \tau_a}{\partial \beta} \quad (6)$$

The equations of the aerosol influence quantifier $\Omega$ for SIMv.2 and REST2 are listed in Appendix B. These equations are more complex than Eq. (6), capturing the second order influence of aerosol specific parameters.

*3.3 Sensitivity of clear-sky models to variations in aerosol loading*

Figure 1 displays the aerosol influence quantifier $\Omega$ as a function of the Ångström turbidity $\beta$ for the models PS, SIMv.2 and REST2. Visual inspection shows a similar qualitative behavior for all models, with a steep rise in $\Omega$ at low turbidity, a maximum, and then a slower asymptotic decrease as $\beta$ grows. For REST2, $\Omega$ lies somewhere between that of SIMv.2 and PS, but its qualitative behavior is more similar to SIMv.2. SIMv.2 and REST2 present a peak at mid-level atmospheric turbidity ($\beta \approx 0.1-0.2$). Although similar in structure, PS experiences a slower increase in $\Omega$ at small $\beta$, and presents a broader plateau than the others, located around $\beta \approx 0.4-0.5$. PS also decreases slower than the other models. In the moderate-to-high turbidity value range

$\beta \in [0.08, 0.5]$, $\Omega$ varies for SIMv.2 between 0.50 and 0.66, for REST2 between 0.54 and 0.64, while for PS between 0.40 and 0.57. We interpret this as PS having a lower sensitivity to changes in aerosols at mid-to-high turbidity.

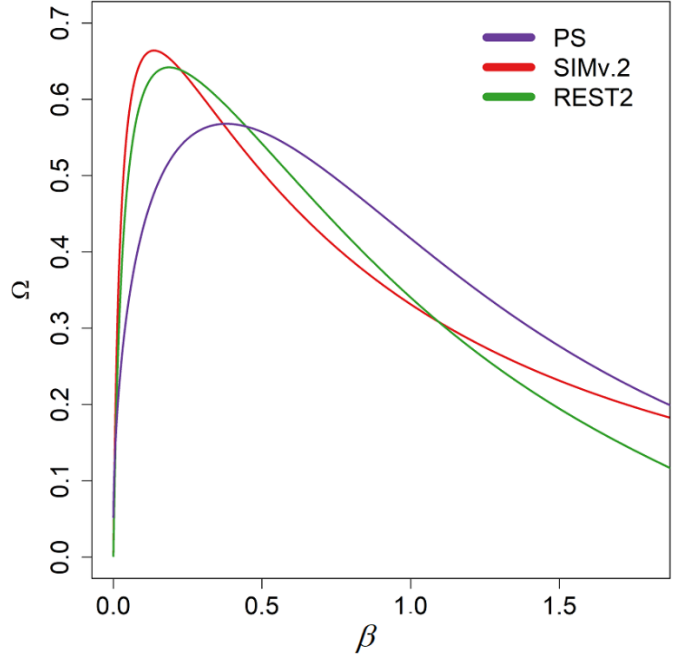

FIGURE 1. The aerosol influence quantifier $\Omega$ estimated by the three clear-sky solar irradiance models (Table 3) as a function of Ångström turbidity coefficient $\beta$. For all other input parameters, the average values over the whole dataset were used (Table 2).

The variation of $\Omega$ with the variation of other atmospheric species is further explored in Fig. 2. Visual inspection shows that the curves for different values of the water vapour column content $w$ overlap exactly for both PS and SIMv.2 (Fig. 2a and Fig. 2b). It confirms that the aerosol influence quantifier $\Omega$ is independent of atmospheric humidity, as the analytically inferred formulas (Eq. (6) and Eq. B2) show. Similarly, in the case of the ozone column content, the curves overlap exactly for PS. $\Omega$ in the case of the PS model depends exclusively on the aerosol loading and the air mass (Eq. 6). The SIMv.2 model uses different ozone transmittances for the diffuse and direct components, which translates into a small difference between the $\Omega$ curves for the maximum and minimum value of $l_{O3}$. SIMv.2 furthermore takes explicitly as input the single scattering albedo $\varpi$ and the aerosol asymmetry factor $g$, which turn out to have a significant impact on $\Omega$. We observe that the variation of $g$ visibly shifts the $\Omega$ curves from the others and each other. The minimum single scattering albedo, $\varpi = 0.407$ in Table 2, changes the overall shape of the graph and the position of the maximum, making the correlation strength more like the one for PS. From a physical point of view, such a low value for $\varpi$ means an aerosol that is more absorbent than scattering and, consequently, the turbidity variation has a smaller influence on the diffuse fraction.

    The situation is similar in the case of REST2 (Fig. 2c). Due to the two-band nature of the model, the slight water vapor impact is visible (because the corresponding specific atmospheric transmittances do not cancel out in the expression of $\Omega$). Similar to SIMv.2, the single-scattering albedo induces a large variation in $\Omega$, and the additional input parameter of REST2, the Ångström exponent $\alpha$ also has a significant impact. A high value of $\alpha$ means aerosol resulted from biomass-burning, expected to be more absorbent. Fig 2c shows that, in terms of Ångström exponent, $\Omega$ captures a similar increase of REST2 sensitivity as the aerosol becomes more scattering. Different from SIMv.2 a decreasing of the single scattering albedo leads to an increase in the diffuse fraction sensibility to the changes in aerosols. Comparing the curves in Fig. 2b and Fig. 2c drawn for the minimum single scattering albedo, the contrast is striking. This proves the capability of the aerosol influence quantifier to bring to light behaviors which the usual tests do not have the ability to reveal.

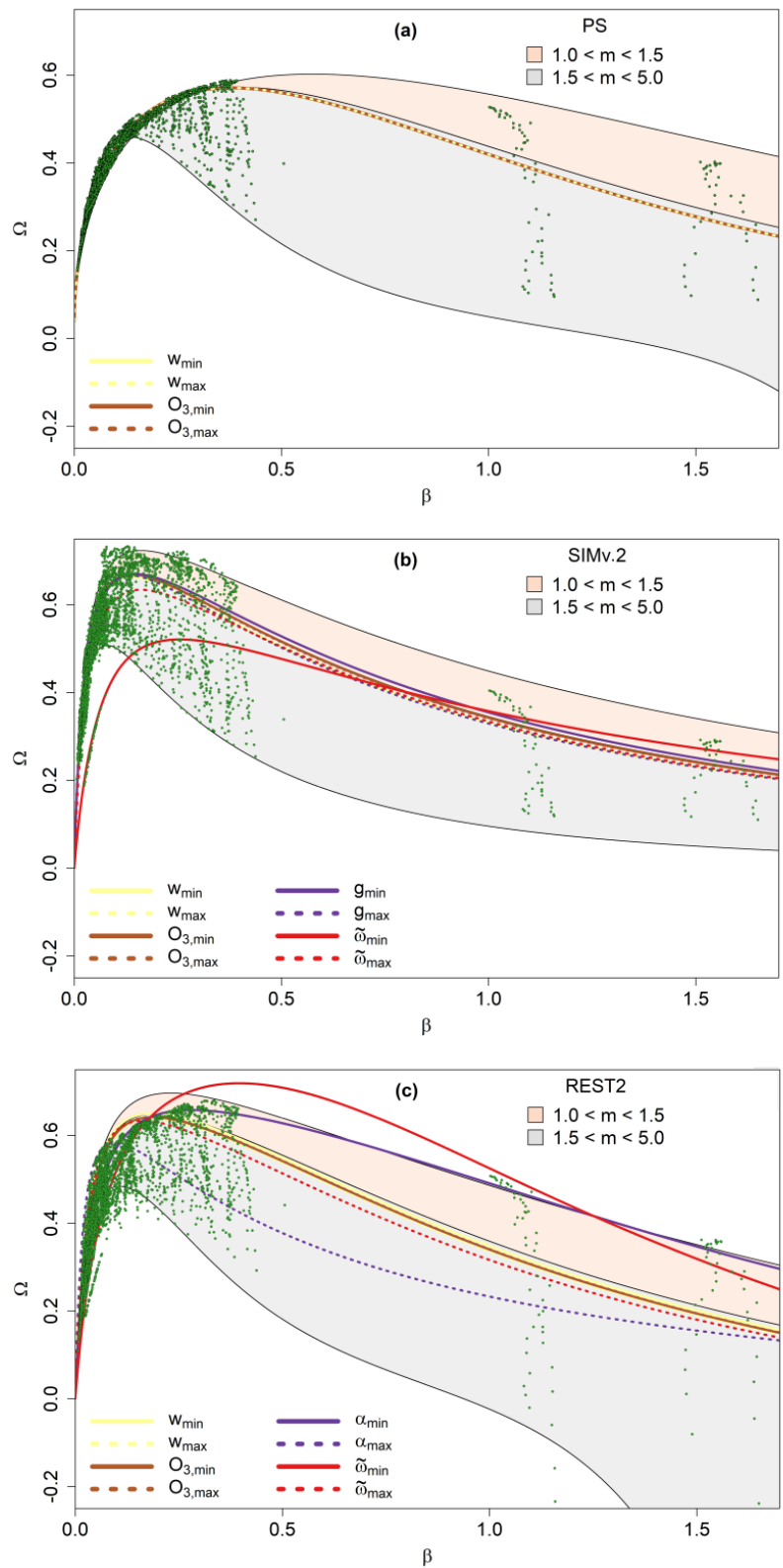

FIGURE 2. Aerosol influence quantifier $\Omega$ estimated by (a) PS, (b) SIM.v2, and (c) REST2 as a function of the Ångström turbidity coefficient $\beta$ for different combinations of the input parameters. The area delimited by the solid lines represents the domain of values taken by $\Omega$ when the atmospheric air mass $m$ varies between 1 (top), 1.5 (middle), and 5 (bottom), with all other input variables taking their average values over the entire dataset (see Table 2). The other curves are evaluated at min./max. value over the entire dataset for one influential parameter and the average value for all others, including the atmospheric air mass ($m = 1.59$). The green dots represent the aerosol influence quantifier evaluated with all measured values from the dataset.

*3.4 Maximum sensitivity*

Irrespective of the clear-sky solar irradiance model, Fig. 1 shows that the aerosol influence quantifier always experiences a maximum value. The physical phenomenology that the three models faithfully capture is the following. As the solar radiation beam passes through the atmosphere, the interactions between the photons in the beam and aerosols result in scattering of the photons out of the beam. The probability of a scattering process basically depends on the aerosol concentration and the photons density in the beam. As the aerosol loading of the atmosphere increases, the probability of scattering photons out of the beam increases too. In their attempt to capture reality, as $\beta$ increases, parametric clear sky models become increasingly sensitive to $\beta$ variation. In other words, as $\beta$ increases, the same variation in $\beta$ will lead to a larger variation in the diffuse fraction. At the same time, along the same path traveled in the atmosphere, the increase in the aerosol loading of the atmosphere leads to a decrease in the density of the solar radiation beam (DNI decreases). This is an opposite process, resulting in decreasing the probability of an aerosol scattering process. At a critical value of the Ångström turbidity coefficient $\beta_c$, the two antagonistic processes are in balance and the clear sky solar irradiance model exhibits maximum sensitivity to the variation of $\beta$. Continuing to increase the aerosol loading of the atmosphere ($\beta > \beta_c$) the concentration of the scattering centers increases too, the solar radiation beam is increasingly rarefied, multiple-scattering processes occur increasing the solar radiation path, all leading to an increasing in the atmosphere opacity. As a natural consequence, the clear sky solar irradiance models sensitivity to the change in the atmospheric aerosol loading decreases too.

The critical value of $\beta$ deserves more attention. Figure 3 displays the aerosol influence quantifier $\Omega$ with respect to the Ångström turbidity factor $\beta$ for different values of atmospheric air mass $m$. Only two clear sky solar irradiance models are considered: PS (Fig. 3a), accounting in the simplest way for the aerosol effects on solar radiation beam, and REST2 (Fig. 3b), accounting in a complex manner for the various facets of aerosol effects. More precisely, the curves from Fig. (3a) and Fig. (3b) are graphical representations of $\Omega(m,\beta)$ given by Eq. (6) and Eq. (B9), respectively. PS was run with the atmospheric optical mass of Kasten and Yang (see the Electronic Supplementary Material) while REST2 was run with the specific atmospheric optical masses from SMARTS2 [23]. In the case of REST2, the atmospheric optical mass associated to the Rayleigh scattering has been considered for building Fig. 3b. For zenithal angle less than 80°, the two equations, Kasten and Yang and that associated to Rayleigh scattering in REST2, provide with enough accuracy the same value for the atmospheric optical mass, further denoted by *m*. In the case of REST2, the remaining input parameters were kept constant at the average values over the entire dataset (Table 2). These values are close to the 45°N climatological values for temperate climate. Thus, the following discussions and equations acquire generality for temperate continental zones. Visual comparison of Figs. (3a) and (3b) shows that the extreme of $\Omega(m,\beta)$ depends differently on the Ångström turbidity coefficient and the atmospheric air mass. For the same range of the air mass, the range of the critical value of the Ångström turbidity factor $\beta_c$ is larger for the PS model than for REST2.

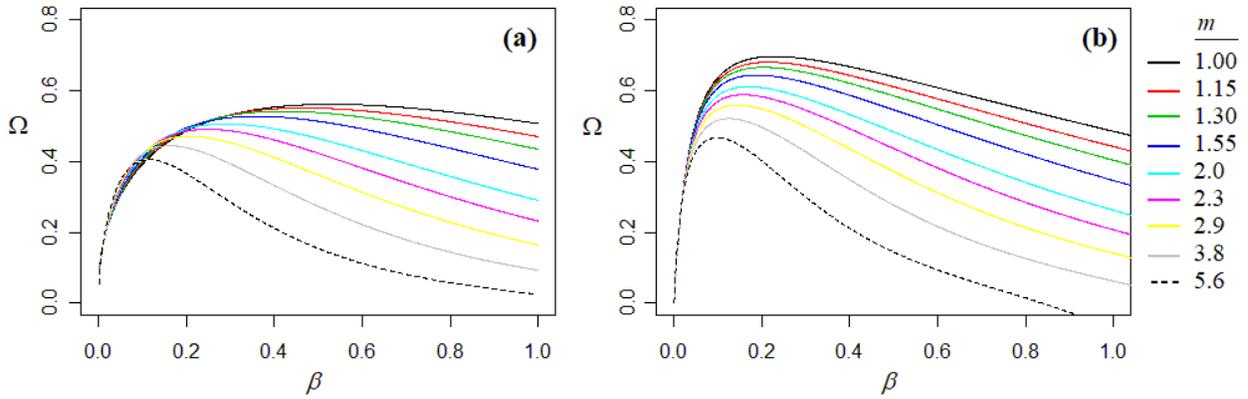

FIGURE 3. Aerosol influence quantifier $\Omega$ with respect to the Ångström turbidity coefficient $\beta$ estimated by two clear sky solar irradiance models: (a) PS (Eq. 6) and (b) REST2 (Eq. B9). The curves are calculated for different atmospheric optical masses *m*.

The magnitude of the maxima $\Omega_{max}(m) = \Omega(m,\beta_c)$ and the relationship between the critical $\beta_c$ and the atmospheric air mass *m* are studied in Fig. 4. Basically, Fig. 4a displays the maxima $\Omega_{max}(m)$ collected from Fig. 3. The graphs portray the observation that the sensitivity of the PS model is always lower than that of REST2.

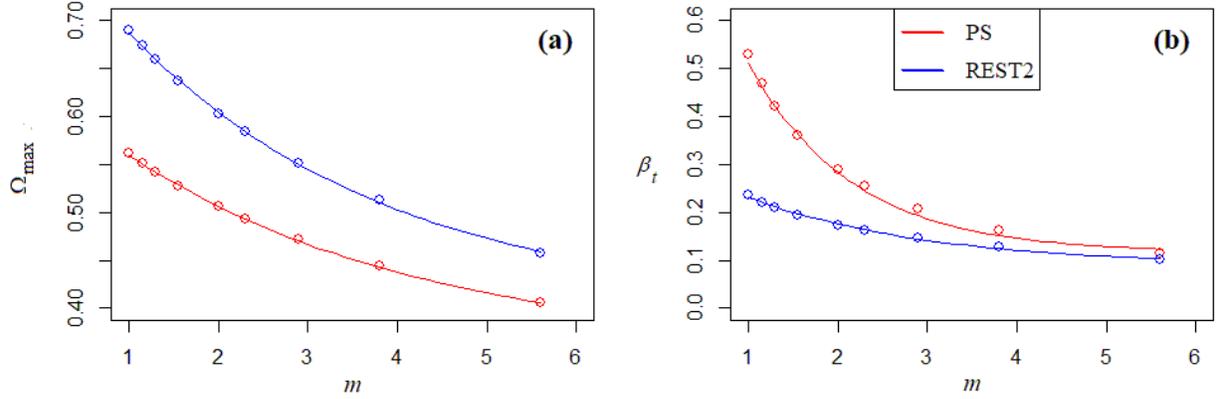

FIGURE 4. (a) Maximum value of the aerosol influence quantifier $\Omega_{max} = \Omega(m, \beta_c)$ with respect to the atmospheric air mass $m$ and (b) The critical value $\beta_c$ of the Ångström turbidity coefficient with respect to the atmospheric air mass $m$.

Considering that in the case of the simple PS model $\Omega = \Omega(m, \beta)$ is a function only of the Ångström turbidity coefficient and the atmospheric air mass, $\Omega_{max}(m)$ defines precisely the maximum sensitivity of the model. Figure 4b shows the dependence of the critical Ångström turbidity coefficient $\beta_c$ to which $\Omega_{max}$ is reached, with respect to the atmospheric air mass $m$. Both graphs can be empirically fitted to exponential equations:

$$\Omega_{max}(m) = 0.356 + 0.276 e^{-0.304 m} \qquad (7a)$$

$$\beta_c(m) = 0.117 + 0.934 e^{-0.864 m} \qquad (7b)$$

for the PS model, and

$$\Omega_{max}(m) = 0.399 + 0.404 e^{-0.341 m} \qquad (8a)$$

$$\beta_c(m) = 0.089 + 0.233 e^{-0.492 m} \qquad (8b)$$

for the REST2 model.

Equations (7) and (8) are of practical importance, because they allow an evaluation of the critical Ångström turbidity coefficient $\beta_c$ and the maximum sensitivity of the models to the changes in the atmospheric aerosol loading based exclusively on the atmospheric air mass. Once such empirical equations are constructed, they can be used in specific applications without any other calculation.

*3.5 Application: variation of clear-sky model accuracy with aerosol temporal resolution*

As Ref. [25] argued, many parametric clear-sky solar irradiance models behave qualitatively similarly to SIM.v2 and REST2 in what regards the sensitivity to changes in $\beta$. The different behavior of the PS model might be the source of its surprising performance in some cases [25]. While this difference in sensitivity is a failure of the model to accurately capture reality, an interesting scenario can arise. Generally, accurate aerosol information is sparse, and climatological values with yearly or monthly granularity are often used in scientific studies. The errors induced by the use of low temporal scale aerosol data on estimated direct normal and global horizontal solar irradiance are well documented (see, for example, [8] and [26]). On the other hand, if high quality aerosol data are not available, one could potentially maximize the accuracy of the results by using a model with low sensitivity to aerosol variations.

Table 4 shows the accuracy— root mean square error normalized to the mean of the measurements *nRMSE*—of global horizontal irradiance estimates using the PS and REST2 models, applied on the working database (Sec. 3.1) and taking $\beta$ at different timescales. REST2 was implemented with a vanishing reflected component. Naturally, as the aerosol timescale increases, the accuracy of both models' output drops drastically. In particular, when changing from daily to monthly timescale, the error in the model estimates doubles for REST2. PS starts out much weaker than REST2 when using instantaneous and daily values. As the timescale increases, the accuracy of PS nears that of REST2, even overtaking it slightly when using monthly values.

Many studies of the impact of aerosols on the collectable solar resource still use monthly aerosol optical depth climatology for modeling [27]. For such long timescales, as we have shown, a simpler model, like PS, can outperform a state-of-the-art model, like REST2. This behavior is caused by the PS less sensitivity to uncertainties introduced using monthly values instead of the actual physical aerosol properties. This feature of PS was anticipated by the aerosol influence quantifier.

TABLE 4. Normalized to the mean root mean square error *nRMSE* [%] at the estimation of global horizontal solar irradiance with the PS and REST2 models. Different time scales for the Ångström turbidity coefficient are used. All other input parameters are used as the measured values from database. Climatologic is defined as the multi-year average at each station.

| Time scale | Instantaneous | Daily | Monthly | Climatologic |
|---|---|---|---|---|
| PS | 20.4 | 23 | 29.8 | 34.7 |
| REST2 | 16.6 | 15.8 | 31 | 33.5 |

## 4. CONCLUSIONS

A substantial uncertainty in modelling global solar irradiance under clear-sky due to inadequate consideration of the aerosol properties is well documented. Moreover, atmospheric aerosol loading can have a strong influence on the dissimilarities between the accuracies in estimating diffuse fraction and the related diffuse solar irradiance [14]. While many studies are concerned with evaluating aerosol influence on the accuracy of solar irradiance estimates, very few studies focus on the aerosol impact on the separation of global solar irradiance into its primary components: direct-normal and diffuse. Aiming to contribute to the latter, this work exploits the capabilities of the clear-sky diffuse fraction in revealing the aerosol impact on the global solar irradiance components. Aerosol influence quantifier $\Omega$ has been introduced as a new measure for the correlation strength between the relative variation in diffuse fraction and the relative variation in atmospheric aerosol loading. $\Omega$ primarily depends on the atmospheric air mass $m$ and the Ångström turbidity coefficient $\beta$. The sensibility of diffuse fraction to finer aerosol properties (e.g. single scattering albedo or aerosol asymmetry factor) is also captured.

In order to illustrate its effectiveness, $\Omega$ was employed to study the sensitivity of three parametric clear-sky solar irradiance models (PS, SIMv.2, and REST2 acknowledged in Sec. 3.2) to variations in aerosol atmospheric loading. For all three models, analytical equations for $\Omega$ are provided. The complex dependencies of $\Omega$ on $m$ and $\beta$ have been brought to light. For a given atmospheric mass, as the atmospheric turbidity increases, the sensibility of the diffuse fraction increases up to a critical value $\beta_c$, after which it decreases. It was found that $\beta_c$ decreases exponentially with the increase in atmospheric mass.

An application is discussed, showing how the performances of PS and REST2 clear-sky solar irradiance models vary with the timescale of aerosol data used as input. A simpler model, like PS, was found to outperform a state-of-the-art model, like REST2, when monthly mean values of the aerosols' properties were used as input. influence quantifierThis surprising result can be explained by the lesser sensitivity of PS to the uncertainties introduced using monthly values instead of the actual physical aerosol properties. Thus, it is shown that the aerosol influence quantifier can be used as a tool to assess the sensitivity of models under various conditions and select an adequate model when high resolution aerosol input data is not available.

ACKNOWLEDGEMENTS

We thank the dedicated personnel for establishing and maintaining the 12 radiometric (BSRN) and photometric (AERONET) stations considered in this study.

APPENDIX

A. *Implementation of $\Omega$ on ground-measured data*

Implementing $\Omega$ requires the evaluation of the differentials in Eq. (3). From a theoretical perspective, the differentiation is done by holding all the other independent variables constant. In practice, the differences can be calculated between any two sets of measurements for which the differences between the relevant parameters are small enough. Strict control of the introduced errors needs to be performed in order to obtain a meaningful result. Considering that the most influential parameters are the aerosols and the air mass (or the solar elevation angle), we searched from the database introduced in Sec. 3.1, data subsets having the same air mass. The aerosol influence quantifier $\Omega$ is then computed linearly between each subsequent pair of measurements.

The resulting values are represented in Fig. A1, for the two most populous air mass bins. No clear pattern for $\Omega$ emerges. The analysis could be homed in further by restricting other parameters, in the order of influence on the extinction of the incoming solar beam. However, further restricting parameter values leads to vanishing bin sizes for our dataset.

The above examples show a clear limitation in attempting to explore $\Omega$ based purely on measurements. We interpret this failure as arising from the limitations of the type of parametrization of the solar irradiance. The broadband solar irradiance values archived in BSRN describe reality at a too coarse level to be able to capture the exact physical impact of small variations in the relevant atmospheric parameters.

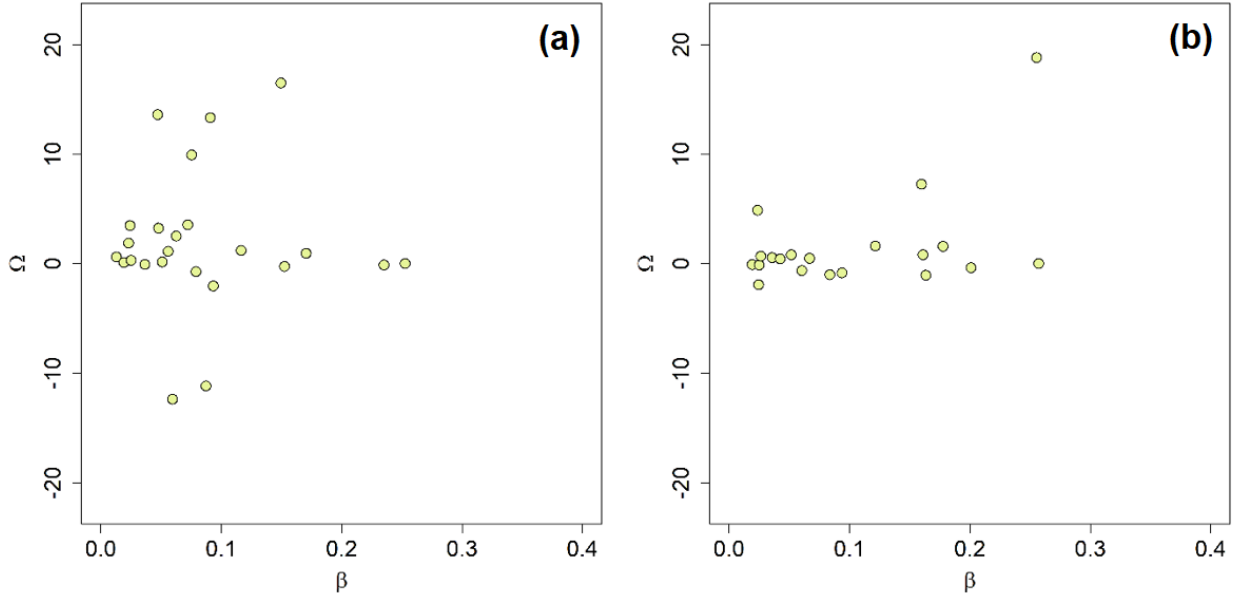

FIGURE A1. Values of the aerosol influence quantifier Ω with respect to the Ångström turbidity coefficient in the two most populous subsets defined by the atmospheric air mass *m*: (a) *m* = 1.07, *N* = 27 and (b) *m* = 1.05, *N* = 20. *N* represents the number of elements in each class.

B. *The aerosol influence quantifier equations for SIMv.2 and REST2*

The equations for the diffuse fraction $k_{d,cs}$ and the aerosol influence quantifier Ω in case of the SIMv.2 and REST2 models are listed below. The equations of the atmospheric transmittances and all other functions that enter the formulas for $k_{d,cs}$ and Ω are presented in the Electronic Supplementary Material.

*SIMv.2*

The clear-sky diffuse fraction is expressed as:

$$k_{d,cs}(m, l_{O3}, \beta, g, \varpi) = 1 - \frac{\tau_{O3}\tau_R\tau_{as}}{\tau_{O3}\tau_R\tau_{as} + \gamma_R\tau_{d,O3}(1-\tau_R)\zeta(\varpi) + \gamma_a\tau_{d,O3}\tau_R(1-\tau_{as})\zeta(\varpi)} \quad (B1)$$

where the specific atmospheric transmittance are denoted as follows: $\tau_{O3}$ for the ozone absorption, $\tau_R$ for the Rayleigh scattering and $\tau_{as}$ for the aerosol scattering process. $\tau_{d,O3}$ denotes the adjusted atmospheric transmittance for ozone extinction in the SIMv.2 diffuse irradiance equation [22]. $\gamma_R$

and $\gamma_a$ represent the downward fractions in the Rayleigh and aerosol scattering processes, respectively.

Simple calculations lead to the equation of aerosol influence quantifier in SIMv.2 model:

$$\Omega(m,l_{O3},\beta,g,\varpi) = \frac{-\beta\tau_{O3}\tau_R\left[\gamma_R(1-\tau_R)+\gamma_a\tau_R\right]}{\left[\gamma_R(1-\tau_R)+\gamma_a\tau_R(1-\tau_{as})\right]\left[\tau_{O3}\tau_R\tau_{as}+\tau_{d,O3}\zeta(\varpi)(\gamma_R(1-\tau_R)+\gamma_a(1-\tau_{as})\tau_R)\right]}\left(\frac{d\tau_{as}}{d\beta}\right)$$

(B2)

where the $\frac{\partial\tau_{as}}{\partial\beta}$ is given in the Electronic Supplementary Material and $\zeta(\varpi)$ denotes:

$$\zeta(\varpi) = \left(0.14578 + \frac{0.69109}{\varpi}\right) \quad (B3)$$

with $\varpi$ representing the single scattering albedo.

## REST2

The clear-sky diffuse fraction is expressed as:

$$k_{d,cs}(m,p,l_{O3},l_{NO2},w,\beta,\alpha,\varpi) = 1 - \frac{P_1+P_2}{P_1+P_2+Q_1+Q_2} \quad (B4)$$

The equations of $P_k$ and $Q_k$ ($k=1,2$ indicating the spectral bands) from Eq. (B4) follow:

$$P_1 = \tau_{R_1}\tau_{g_1}\tau_{o_1}\tau_{n_1}\tau_{w_1}\tau_{a_1}G_{0n_1} \quad (B5)$$

$$P_2 = \tau_{R_2}\tau_{g_2}\tau_{w_2}\tau_{a_2}G_{0n2} \quad (B6)$$

$$Q_1 = \tau_{g_1}\tau_{o_1}\tau'_{n_1}\tau'_{w_1}\left[B_{R_1}(1-\tau_{R_1})\tau_{a_1}^{0.25}+B_aF_1\tau_{R_1}(1-\tau_{a_1}^{0.25\omega_1})\right]G_{0n_1} \quad (B7)$$

$$Q_2 = \tau_{g_2}\tau'_{w_2}\left[B_{R_2}(1-\tau_{R_2})\tau_{a_2}^{0.25}+B_aF_2\tau_{R_2}(1-\tau_{a_2}^{0.25\omega_2})\right]G_{0n_2} \quad (B8)$$

The terms $P_k$ are the direct normal irradiance in the band $k$. $\tau_{R_k}$, $\tau_{g_k}$, $\tau_{O_k}$, $\tau_{n_k}$, $\tau_{w_k}$, $\tau_{a_k}$ are the band transmittance functions for Rayleigh scattering, uniformly mixed gases absorption, ozone absorption, nitrogen dioxide absorption, water vapor absorption and aerosol extinction, respectively. The terms $Q_k \cdot \cos\theta_z$ are the incident diffuse irradiance on a perfectly absorbing ground in the band $k$. $\tau'_{n_k}$ and $\tau'_{w_k}$ are transmittances for nitrogen dioxide and water vapor down scattering in the bottom layer. $F_k$ is a correction factor introduced to compensate for multiple scattering effects. $B_{R_k}$ are the forward scattering fractions for Rayleigh extinction and $B_a$ is aerosol forward scattering factor. $G_{0n_k}$ are the extra-atmospheric irradiances at the mean sun-earth distance.

We remind the reader that the explicit equations for all these atmospheric transmittances, correction factors and scattering fractions are reproduced in the Electronic Supplementary Material.

Simple but laborious calculations lead to:

$$\Omega(m,\beta,p,l_{O3},l_{NO2},w,\alpha,\varpi) = \frac{\beta\left[(P_1+P_2)\left(\frac{\partial Q_1}{\partial \beta}+\frac{\partial Q_2}{\partial \beta}\right)-(Q_1+Q_2)\left(\frac{\partial P_1}{\partial \beta}+\frac{\partial P_2}{\partial \beta}\right)\right]}{(Q_1+Q_2)(P_1+P_2+Q_1+Q_2)} \quad (B9)$$

The derivatives for the terms $P_k$ and $Q_k$ ($k=1,2$) are:

$$\frac{\partial P_1}{\partial \beta} = \tau_{R_1}\tau_{g_1}\tau_{o_1}\tau_{n_1}\tau_{w_1}G_{0n_1}\frac{\partial \tau_{a_1}}{\partial \beta} \quad (B10)$$

$$\frac{\partial P_2}{\partial \beta} = \tau_{R_2}\tau_{g_2}\tau_{w_2}G_{0n2}\frac{\partial \tau_{a_2}}{\partial \beta} \quad (B11)$$

$$\frac{\partial Q_1}{\partial \beta} = \tau_{g_1}\tau_{o_1}\tau'_{n_1}\tau'_{w_1}\left[\frac{1}{4}B_{R_1}(1-\tau_{R_1})\tau_{a_1}^{-\frac{3}{4}}\frac{\partial \tau_{a_1}}{\partial \beta}+B_a\frac{\partial F_1}{\partial \beta}\tau_{R_1}\left(1-\tau_{a_1}^{\frac{\varpi_1}{4}}\right)-\frac{1}{4}\varpi_1 B_a F_1 \tau_{R_1}\tau_{a_1}^{\frac{\varpi_1}{4}-1}\frac{\partial \tau_{a_1}}{\partial \beta}\right]G_{0n_1} \quad (B12)$$

$$\frac{\partial Q_2}{\partial \beta} = \tau_{g_2}\tau'_{w_2}\left[\frac{1}{4}B_{R_2}(1-\tau_{R_2})\tau_{a_2}^{-\frac{3}{4}}\frac{\partial \tau_{a_2}}{\partial \beta}+B_a\frac{\partial F_2}{\partial \beta}\tau_{R_2}\left(1-\tau_{a_2}^{\frac{\varpi_2}{4}}\right)-\frac{1}{4}\varpi_2 B_a F_2 \tau_{R_2}\tau_{a_2}^{\frac{\varpi_2}{4}-1}\frac{\partial \tau_{a_2}}{\partial \beta}\right]G_{0n_2} \quad (B13)$$

The derivatives $\frac{\partial \tau_{a_1}}{\partial \beta}$, $\frac{\partial \tau_{a_2}}{\partial \beta}$, $\frac{\partial F_1}{\partial \beta}$ and $\frac{\partial F_2}{\partial \beta}$ are given in Electronic Supplementary Material.